\documentclass[12pt]{article}
\usepackage{setspace}
\usepackage{graphicx}
\date{}
\def\b{\begin{equation}}
\def\e{\end{equation}}
\def\bee{\begin{enumerate}}
\def\eee{\end{enumerate}}
\title{Bell's inequality with Dirac particles }
\author{Shahpoor Moradi$^1$,
 \thanks{e-mail: shahpoor.moradi@gmail.com} }
\date{\today}

\begin{document}
\maketitle {\it \centerline{$^1$  Department of Physics, Razi
University, Kermanshah, IRAN}}
 \maketitle {\it \centerline{\emph{Tel: +989183360186, Fax:
 +988314275599}}}

\begin{abstract}
We study Bell's inequality using the Bell states constructed from
four component Dirac spinors. Spin operator is related to the
Pauli-Lubanski pseudo vector which is relativistic invariant
operator. By using Lorentz transformation, in both Bell states and
spin operator, we obtain an observer independent Bell's
inequality, so that it is maximally violated as long as it is
violated maximally in the rest frame.
\end{abstract}

 \vspace{0.2cm}
{\it \emph{Proposed PACS number}}: \emph{03.65.Ud} \vspace{0.5cm}

\vspace{0.5cm} {\textit{Keywords}}: {Bell's inequality, Lorentz
transformation}

\newpage Relativistic entanglement and
quantum nonlocality  is investigated by many authors [1-28]. M.
Czachor \cite{cz}, investigated Einstein-Podolsky-Rosen experiment
with relativistic massive spin-$\frac{1}{2}$ particles. The degree
of violation of the Bell's inequality is shown to depend on the
velocity of the pair of  particles with respect to the laboratory.
 He considered the spin singlet of two spin-$\frac{1}{2}$ massive particles moving in the
same direction. He introduced the concept of a relativistic spin
observable using the relativistic center-of-mass operator. For two
observers in the lab frame measuring the spin component of each
particle in the same direction, the expectation value of the joint
spin measurement, i.e., the expectation value of the tensor
product of the relativistic spin observable of each constituent
particle, depends on the boost velocity. Only when the boost speed
reaches that of light, or when the direction of the spin
measurements is perpendicular to the boost direction, the results
seem to agree with the EPR correlation. In the present article we
use a relativistic spin operator for spin-$\frac{1}{2}$ particles
which is constructed from Pauli-Lubanski pseudo vector. Also we
use the four components Dirac spinors for constructing Bell
states. With this spin operator the Bell's inequality is maximally
violated and Bell observable does not depend on velocity of
particles.

The Pauli-Lubanski pseudo vector is \b
W^{\mu}=\frac{i}{4}\epsilon^{\mu\nu\rho\sigma}\sigma_{\nu\rho}\partial_{\sigma}
,\e where $\sigma_{\mu\nu}=\frac{i}{2}[\gamma_{\mu},\gamma_{\nu}]$
and in the Dirac representation the $\gamma$-matrices are \b
\gamma^0=\left(%
\begin{array}{cc}
  {\mathbf{1}} & 0 \\
 0 &  -{\mathbf{1}} \\
\end{array}%
\right),\;\;\;\;{{\gamma}}^i=\left(%
\begin{array}{cc}
  0 & {\sigma}^i \\
-\mathbf{\sigma}^i &  0 \\
\end{array}%
\right),\;\;\;\;\gamma^5=\left(%
\begin{array}{cc}
  0 & {\mathbf{1}} \\
{\mathbf{1}} &  0 \\
\end{array}%
\right).\e  The invariant  spin observable for Dirac particle is
\b
\hat{s}=\frac{2W_{\mu}s^{\mu}}{m}=\pm\frac{1}{m}\gamma_5{\not{s}}{\not{p}}=\gamma_5{\not{s}},\e
where $s^{\mu}$ is space-like normalized four-vector orthogonal to
${\mathbf{p}}$. The pervious equation holds for plane wave
solutions, upper(lower) sign is related to positive(negative)
energy solutions. For rest particles the vector $s^{\mu}$ becomes
$(0,{\mathbf{n}})$, so that \b
\hat{s}=\gamma_5\gamma_0{\mathbf{n}}\cdot\gamma={\mathbf{\Sigma}}\cdot{\mathbf{n}}
.\e If we choose $s$ along z axis, we see that following spinors
are eigenstates of $\Sigma_z$ \b
u({\mathbf{0}},\frac{1}{2})=\left(%
\begin{array}{c}
  1 \\
  0 \\
  0 \\
  0 \\
\end{array}%
\right),\;\;\;\;\;
u({\mathbf{0}},-\frac{1}{2})=\left(%
\begin{array}{c}
  0 \\
  1 \\
  0 \\
  0 \\
\end{array}%
\right), \e \b
v({\mathbf{0}},\frac{1}{2})=\left(%
\begin{array}{c}
  0 \\
  0 \\
  1 \\
  0 \\
\end{array}%
\right),\;\;\;\;\;
v({\mathbf{0}},-\frac{1}{2})=\left(%
\begin{array}{c}
  0 \\
  0 \\
  0 \\
  1 \\
\end{array}%
\right), \e
 with eigenvalues
$+1$ for $u({\mathbf{0}},\frac{1}{2})$ and
$v({\mathbf{0}},\frac{1}{2})$, and $-1$ for
$u({\mathbf{0}},-\frac{1}{2})$ and $v({\mathbf{0}},-\frac{1}{2})$.
Entangled  Bell state corresponding to positive energy spinors
$u({\mathbf{0}},\pm\frac{1}{2})$ is defined as \b
\Phi_{11}=\frac{1}{\sqrt{2}}\left(u({\mathbf{0}},\frac{1}{2})\otimes
u({\mathbf{0}},-\frac{1}{2})-u({\mathbf{0}},-\frac{1}{2})\otimes
u({\mathbf{0}},\frac{1}{2})\right). \e The two particle states
involve only positive energy spinors $u$ and not the negative
energy spinors $v$. This occurs because the positive energy states
transformed among themselves separately and do not mix with each
other under Lorentz transformation \cite{am}. We consider
measurements of the spins ${\mathbf{\Sigma}}_1$ and
${\mathbf{\Sigma}}_2$ along different directions performed on the
correlated particles 1 and 2. For particle 1, we measure either
$\hat{a}$ or $\hat{a}'$ where $\hat{a} =
{\mathbf{a}}\cdot{\mathbf{\Sigma}}_1$ and $\hat{a}' =
{\mathbf{a}}'\cdot{\mathbf{\Sigma}}_1$ . Measured values of
$\hat{a}$, $\hat{a}'$ are $\pm 1$. Similarly, for particle 2, the
quantity $\hat{b}$ or $\hat{b}'$ is measured, where
$\hat{b}={\mathbf{b}}\cdot{\mathbf{\Sigma}}_2$ and $\hat{b}' =
\mathbf{b}'\cdot{\mathbf{\Sigma}}_2$. One of the form of Bell's
inequality is  \b |\langle \hat{a}\otimes\hat{b}\rangle +\langle
\hat{a}'\otimes\hat{b}\rangle + \langle
\hat{a}\otimes\hat{b}'\rangle-\langle
\hat{a}'\otimes\hat{b}'\rangle|\leq 2.\e This inequality is
violated by the quantum mechanical results for, say, the singlet
state (7) where
 $ \langle \hat{a}\otimes\hat{b}\rangle=-{\mathbf{a}}\cdot{\mathbf{b}}$. It's easy to check that
  the maximally violation of  Bell's inequality
 is equal to $2\sqrt{2}$.

 To construct the spinors in an arbitrary Lorentz
frame we boost the rest frame spinors by the standard  boost from
the rest frame to a frame with momentum ${\mathbf{p}}$. Then the
 entangled Bell states (7) takes the form \b
\Psi_{11}=\frac{1}{\sqrt{2}}\left(u({\mathbf{p}},\frac{1}{2})\otimes
u({\mathbf{p}},-\frac{1}{2})-u({\mathbf{p}},-\frac{1}{2})\otimes
u({\mathbf{p}},\frac{1}{2})\right).\e Where the normalized
positive energy spinors  $u({\mathbf{p}},\pm\frac{1}{2}) $ are
 \b
u({\mathbf{p}},\pm\frac{1}{2})
=\sqrt{\frac{E_p+m}{2E_p}}\left(\begin{array}{c}
  \varphi^{(\pm)} \\
 \frac{\sigma\cdot{\mathbf{p}}}{E_p+m} \varphi^{(\pm)}
\end{array}\right),\e
with $\sigma_3\varphi^{(\pm)}=\pm\varphi^{(\pm)}$. Now the Lorentz
transformed spinors $u({\mathbf{p}},\pm\frac{1}{2}) $ are
eigenstates of $\gamma_5{\not{s}}$ where $s$ is the transform of
$s_z$. In the frame in which the electron has momentum
${\mathbf{p}}$ the polarization vector is obtained by applying the
Lorentz boost \b
s^{\mu}=(s_0,{\mathbf{s}})=\left(\frac{{\mathbf{p}\cdot\mathbf{n}}}{m},{\mathbf{n}}+
\frac{({\mathbf{p}\cdot\mathbf{n}}){\mathbf{p}}}{m(E_p+m)}\right)
.\e  Expectation value of the spin projection  $\gamma_5{\not{s}}$
in state $u({\mathbf{p}})$ is \b
\langle\gamma_5{\not{s}}\rangle=\frac{E_p+m}{2E_p}
\left(\varphi^{\dag}({\mathbf{\sigma}}\cdot{\mathbf{s}})\varphi-
\frac{\varphi^{\dag}({\mathbf{\sigma}}\cdot{\mathbf{p}})
({\mathbf{\sigma}}\cdot{\mathbf{s}})({\mathbf{\sigma}}\cdot{\mathbf{p}})\varphi}{(E_p+m)^2}\right).
\e Using the following identity
 \b
({\mathbf{\sigma}}\cdot{\mathbf{p}})
({\mathbf{\sigma}}\cdot{\mathbf{s}})({\mathbf{\sigma}}\cdot{\mathbf{p}})=|{\mathbf{p}}|^2
(s_z\sigma_z-s_y\sigma_y-s_x\sigma_x), \e
 we have  \b
u^{\dag}({\mathbf{p}},\frac{1}{2})(\gamma_5{\not{s}})u({\mathbf{p}},\frac{1}{2})=
\gamma^{-1}s_z,
 \e
\b
u^{\dag}({\mathbf{p}},-\frac{1}{2})(\gamma_5{\not{s}})u({\mathbf{p}},-\frac{1}{2})=-
\gamma^{-1}{s_z},
 \e
\b
u^{\dag}({\mathbf{p}},\frac{1}{2})(\gamma_5{\not{s}})u({\mathbf{p}},-\frac{1}{2})=s_x-is_y,
 \e
\b u^{\dag}({\mathbf{p}},-\frac{1}{2})(\gamma_5{\not
{s}})u({\mathbf{p}},\frac{1}{2})=s_x+is_y,
 \e
where $\gamma=E/m=(1-u^2)^{-1/2}$. Without loss of generality we
assume that ${\mathbf{p}}=p\hat{z}$, then after some algebra the
average of Bell operator on state $\Psi_{11}$ to be \b
\langle{\Psi_{11}}|{\hat{a}}\otimes{\hat{b}}|{\Psi_{11}}\rangle=
-{\mathbf{a}}\cdot{\mathbf{b}},\e which results the Lorentz
invariant Bell's inequality.  Here we compare our results with
Czachor work [7]. He considered to spin singlet of two
spin-$\frac{1}{2}$ massive particles moving in the same direction.
The normalized operator corresponding to the spin projection along
arbitrary direction ${\mathbf{a}}({\mathbf{a}}^2=1)$ is \cite{cz}
\b\hat{a}=\frac{(\sqrt{1-\beta^2}{\mathbf{a}}_\bot+{\mathbf{a}}_\|).{\mathbf{\sigma}}}
{\sqrt{1-({\mathbf{a}}\times{\mathbf{u}}})^2},\e

 where the
subscripts $\bot$ and $\|$ denote the components which are
perpendicular and parallel to the boost speed ${\mathbf{u}}$. In
this case the average of the relativistic EPR operator to be\b
\langle{\Psi_{11}}|{\hat{a}}\otimes{\hat{b}}|{\Psi_{11}}\rangle=-
\frac{{\mathbf{a}}\cdot{\mathbf{b}}-u^2{\mathbf{a}}_\bot\cdot{\mathbf{b}}_\bot}
{{\sqrt{1-({\mathbf{a}}\times{\mathbf{u}}})^2}{\sqrt{1-({\mathbf{b}}\times{\mathbf{u}}})^2}},\e
Then against our results the expectation value is not Lorentz
invariant and consequently the Bell's inequality is speed
dependent.

In conclusion we have discussed EPR-type experiment with Dirac
particles. For obtaining relativistic invariant Bell's inequality
the spin operator for Dirac particles should be Lorentz
transformed. Then in opposition of Czachor's results\cite{cz},
since the degree of violation does not depend on momentums of
particles, we have the relativistic invariant Bell's inequality.

\bibliographystyle{plain}

\end{document}